# A Preliminary Study of Machine-Learning-Based Ranging with LTE Channel Impulse Response in Multipath Environment

Halim Lee and Jiwon Seo
*School of Integrated Technology, Yonsei University, Korea*
{halim.lee, jiwon.seo}@yonsei.ac.kr

**Abstract**

*Alternative navigation technology to global navigation satellite systems (GNSSs) is required for unmanned ground vehicles (UGVs) in multipath environments (such as urban areas). In urban areas, long-term evolution (LTE) signals can be received ubiquitously at high power without any additional infrastructure. We present a machine learning approach to estimate the range between the LTE base station and UGV based on the LTE channel impulse response (CIR). The CIR, which includes information of signal attenuation from the channel, was extracted from the LTE physical layer using a software-defined radio (SDR). We designed a convolutional neural network (CNN) that estimates ranges with the CIR as input. The proposed method demonstrated better ranging performance than a received signal strength indicator (RSSI)-based method during our field test.*

## 1. Introduction

The global navigation satellite systems (GNSSs) are widely used for determining absolute positions of users [1-3]. Although GNSSs have superior positioning performance, several limitations also exist, which leads to the necessity of alternative or complementary navigation technologies [4-6]. GNSS signals are susceptible to radio interferences such as jamming [7-9], spoofing [10, 11], and ionospheric disturbances [12-14]. Further, GNSS accuracy and availability can be highly degraded in urban areas, indoors, and tunnels.

Among alternative navigation technologies [15-19], long-term evolution (LTE) signal-based navigation has the advantage of utilizing existing infrastructure. In addition, the downlink LTE signals are freely available for navigation purposes. To estimate the absolute position of a user using LTE signals, several types of positioning methods (e.g., time-of-arrival (TOA) [20, 21], time-difference-of-arrival (TDOA) [22, 23], received signal strength indicator (RSSI) [24], and channel impulse response (CIR) [25-27] based methods) have been developed. Among those positioning methods, the RSSI-based and CIR-based positioning methods have the advantage of low complexity. Both RSSI and CIR indicate the channel link quality, but CIR is more attractive for positioning since it contains more information about the channel than RSSI because CIR is estimated from a wider bandwidth and narrower time interval. Recent studies on fingerprinting-based localization [28, 29] using CIR or channel frequency response (CFR), which is the Fourier transform of the CIR, showed higher positioning performance than the cases using RSSI.

Several studies on fingerprinting-based localization using CIR or CFR of LTE signals have been conducted [25-27]. In [25], the descriptors such as mean, spectral slope, and spectral moment which describe the shape of CFR were suggested as fingerprinting metrics. In [26], a neural network-based fingerprinting method using live LTE signals was presented. However, significant time and efforts are required to create fingerprinting maps. Considering the difficulty of LTE signal surveying for fingerprinting map generation, it can be more practical to estimate the receiver's position based on the estimated ranges between the receiver and base stations.

In this paper, we developed a machine-learning-based ranging method to estimate the range between the base station and the receiver by using the CIR of LTE signals. CIR was extracted from live LTE signals using software-defined radio (SDR). LTE signals collected at multiple signal reception spots were used to train the convolutional neural network (CNN), which was designed to estimate the range from CIR.

## 2. Proposed Method

### 2.1 Channel Impulse Response

The received LTE signal in the frequency domain can be modeled as follows [30].

$$Y = CFR \cdot X + N, \tag{1}$$

where $Y$ is the received signal, $CFR$ is the channel frequency response, $X$ is the transmitted signal, and $N$ is noise.

The CFR is obtained by following the procedures in Fig. 1. After receiving the LTE signal, the receiver first removes the frequency offset of the time-domain waveform. Then, the receiver identifies the cell identity and frame start timing through a cell search. After the incomplete frame removal and OFDM demodulation, pilot-symbol-based channel estimation is performed in the frequency domain.

The CRS is assigned to grids at specific positions of the LTE frame according to the antenna port number, as shown in Fig. 2. In this study, the CFR inferred from the CRS that is related to the antenna port 0 for a normal cyclic prefix was utilized. The CFR can be inferred from the CRS of the LTE physical layer because the sequence of the CRS is known to the receiver. For a known sequence, the estimated CFR can be expressed as follows [30].

$$\widetilde{CFR} = \frac{Y}{X_{known}} + N, \qquad (2)$$

The effect of noise can be minimized using the least-squares method.

Fig. 3 presents the CIR plots for a line-of-sight (LOS) signal and an NLOS and severe multipath signal. The horizontal and vertical axes of the CIR plot are the time and symbol index, respectively. The CIR at time $t$ is expressed as in Eq. (3) [30].

$$CIR(t) = \sum_{l=0}^{L-1} |CIR(l)| \cdot \delta(t - t(l)), \qquad (3)$$

where $t$ is the time, $l$ is the index of the impulse, and $L$ is the number of impulses.

The magnitude of the CIR represents the amount of signal attenuation by the channel and can be represented as a two-dimensional image as in Fig. 3. Thus, it can be used as an input to a CNN model.

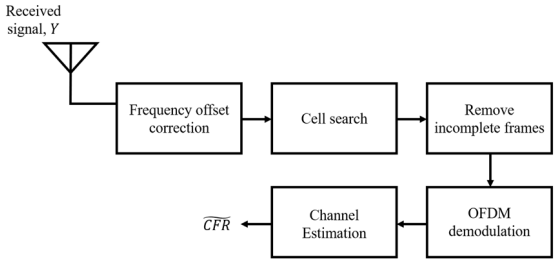

Fig. 1 LTE channel estimation process.

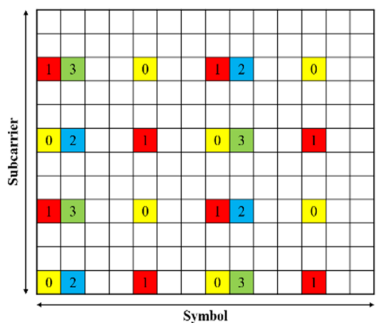

Fig. 2 Positions of the CRS on the LTE frame according to the antenna port for a normal cyclic prefix. The numbers indicated are antenna port numbers, and the port 0 in yellow is used in this study (reproduction of Fig. 1 in [25]).

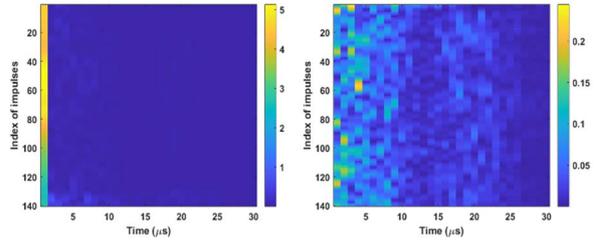

Fig. 3 CIR for a LOS signal (left) and an NLOS and severe multipath signal (right).

## 2.2 CNN model

We designed a CNN structure in which the magnitude of CIR is processed through. For feature extraction, three convolutional layers and three max-pooling layers were used. The number of filters in the first and second convolutional layers is 32 and that in the third convolutional layer is 64. The filter size of all convolutional layers is 3×3. In addition, the kernel size of the max-pooling layer is 2×2. The last max-pooling layer is followed by one flatten layer and two fully connected layers. A rectified linear unit (ReLU) was used as the activation function. The output of the last activation function is the estimated range. The CNN was implemented on TensorFlow. The batch size was set to 256. Moreover, an AdamOptimizer was used to minimize the mean squared error of the ranging.

## 3. Experimental Setup

In the experiment, a UGV was equipped with an ADRV9361-Z7035 SDR 2×2 system on module and ADRV1CRR-FMC carrier module of Analog Devices, commonly collectively referred to as PicoZed software-defined radio (SDR). In addition, the PicoZed SDR was connected to a WA700/2700 antenna from PulseLarsen Antennas, which can receive LTE signals from 698 MHz to 2.7 GHz. The ground truth trajectory of the UGV was obtained using a GNSS real-time kinematic (RTK) receiver, which provides position accuracy at the centimeter level. Fig. 4 shows the equipped UGV.

Band 5 LTE signals with a center frequency of 889 MHz and bandwidth of 10 MHz were received. The location and cell identity of the base station were assumed to be known to the receiver through a database. The locations of the LTE base station (eNodeB) and signal reception spots are shown in Fig. 5. We collected 100–200 frames of LTE signals at 39 spots and the frames were stored and postprocessed.

Two scenarios are considered. The true range between the UGV and base station is in the range of 60–100 m for the first scenario and 100–200 m for the second scenario. The first scenario corresponds to a line-of-sight (LOS) environment, and the second scenario corresponds to a severe multipath environment. For the first scenario, there are 21 training spots marked with red dots and 6 test spots marked with green dots, as shown in Fig. 5(a). In addition, for the second scenario, there are 12 training spots and 3 test spots as shown in Fig. 5(c). The CIRs collected at the training spots were used as the input of the proposed neural network. The ranging performance was evaluated at the test spots.

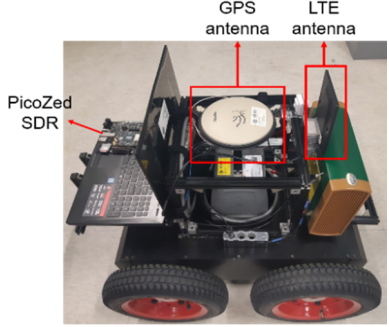

Fig. 4 UGV equipped with a GNSS RTK receiver, LTE antenna, and PicoZed SDR.

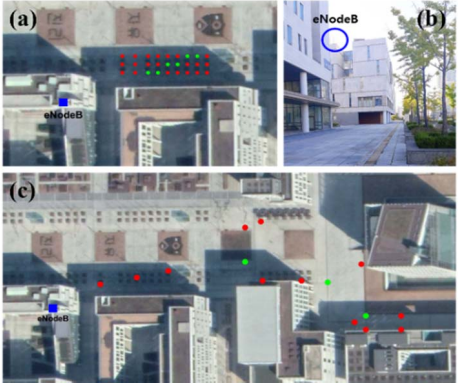

Fig. 5 Experiment spots for (a) line-of-sight and (c) multipath scenarios. Red dots indicate the training spots and green dots indicate test spots. The base station is shown in (b).

## 4. Experimental Results

We compared the performance of our proposed CIR-based ranging method with an RSSI-based ranging method. We designed a different neural network for the RSSI-based ranging because RSSI is one-dimensional data; hence, the proposed neural network for two-dimensional data is not suitable. The neural network for RSSI-based ranging consists of two fully connected layers and a ReLU function. When comparing the neural network for the RSSI-based ranging and proposed CIR-based ranging, the only difference is the convolutional layers for feature extraction. The fully connected layers and activation function for range estimation are the same for both the methods.

Tables 1 and 2 present the means (i.e., biases) and standard deviations of the proposed CIR-based ranging and RSSI-based ranging errors for the first and second scenarios, respectively. Our proposed CIR-based ranging demonstrated better performance than the RSSI-based ranging.

Table 1 Bias and standard deviation of the ranging error for the first scenario.

|  | Bias (m) | Std. Dev. (m) |
| --- | --- | --- |
| RSSI-based ranging | 2.82 | 8.86 |
| CIR-based ranging | 0.71 | 6.17 |

Table 2 Bias and standard deviation of the ranging error for the second scenario.

|  | Bias (m) | Std. Dev. (m) |
| --- | --- | --- |
| RSSI-based ranging | 2.96 | 12.29 |
| CIR-based ranging | 0.76 | 11.94 |

## 5. Conclusion

We developed a method to estimate the range between a UGV and an LTE base station using the CIR extracted from the LTE physical layer. A neural network utilizing the magnitude of CIR was designed. The performance of the proposed method was evaluated under the LOS and multipath scenarios. In both the scenarios, the proposed method demonstrated better performance than the RSSI-based raging method.

## Acknowledgement

This research was supported by the Ministry of Science and ICT (MSIT), Korea, under the High-Potential Individuals Global Training Program (2020-0-01531) supervised by the Institute for Information & Communications Technology Planning & Evaluation (IITP).